\documentclass[11pt, a4paper]{article}%%%%where rsproca is the template name
\usepackage[T1]{fontenc}
\usepackage[utf8]{inputenc}
\usepackage{microtype}
\usepackage{lmodern}
\usepackage[english]{babel}
\usepackage[titletoc]{appendix}
\usepackage{upgreek}

\usepackage{amscd}
\usepackage{amsmath}
\usepackage{amsfonts}
\usepackage{amssymb}
\usepackage{amsthm}
\usepackage{braket}
\usepackage{bm}
\usepackage[mathcal]{euscript}
\usepackage{mathrsfs}
\usepackage{mathtools}
\usepackage{tensor}
\usepackage{enumerate}

\usepackage{hyperref}
\usepackage{bookmark}
\usepackage{enumitem}

\title{Interfacial waveforms in chiral lattices with gyroscopic spinners}
\author{M. Garau\footnote{Keele University, School of Computing and Mathematics, Keele, ST5 5BG, UK}, G. Carta\footnote{Liverpool John Moores University, Mechanical Engineering and Materials Research Centre, Liverpool, L3 3AF, UK},  M.J. Nieves\footnote{University of Cagliari, Department of Mechanical, Chemical and Material Engineering, Cagliari, 09123, Italy. Corresponding author, email: m.nieves@keele.ac.uk}\,\,$^,$\footnotemark[1],
 I.S. Jones\footnotemark[2], \\
N.V. Movchan\footnote{University of Liverpool, Department of Mathematical Sciences, Liverpool, L69 7ZL, UK}\, and A.B. Movchan\footnotemark[4]}

\date{}
\begin{document}

\maketitle

%\address{$^{1}$Keele University, School of Computing and Mathematics, Keele, ST5 5BG, UK\\
%$^{2}$Liverpool John Moores University, Mechanical Engineering and Materials Research Centre, Liverpool, L3 3AF, UK\\
%$^{3}$University of Cagliari, Department of Mechanical, Chemical and Material Engineering, Cagliari, 09123, Italy\\
%$^{4}$University of Liverpool, Department of Mathematical Sciences, Liverpool, L69 7ZL, UK}

%\subject{wave motion, mechanics, mathematical physics}

%\keywords{elastic lattices, gyroscopic spinners, chiral systems, dispersion, uni-directional waveforms}

%\corres{M.J. Nieves\\
%\email{m.nieves@keele.ac.uk}}

\begin{abstract}
We demonstrate a new method of achieving topologically protected states in a discrete hexagonal lattice by attaching gyroscopic spinners, which bring chirality to the system.
Dispersive features of this medium are investigated in detail and, by tuning the parameters of the spinners, it is shown one can manipulate the locations of stop-bands and Dirac points.
We show that in proximity of such points, uni-directional interfacial waveforms can be created in an inhomogeneous lattice and the direction of such waveforms can be controlled.
The effect of inserting additional internal links into the system, which is thus transformed in a heterogeneous triangular lattice, is also investigated.
This work introduces a new perspective in the design of periodic media possessing non-trivial topological features.
\end{abstract}


%%%%%%%%%% Insert the texts which can accomodate on firstpage in the tag "fmtext" %%%%%
%\begin{fmtext}

\section{Introduction}
\label{Intro}

Systems supporting topologically protected edge modes have attracted increasing attention  in recent years. In such systems, which are collectively known as ``topological insulators'',  waves propagating along the edges of the domain are not scattered backwards or inside the medium,
%\newline
%\end{fmtext}
%%%%%%%%%%%%%%% End of first page %%%%%%%%%%%%%%%%%%%%%
%\noindent
even in the presence of imperfections or discontinuities (such as sharp corners and localised defects).  Topological insulators are generally characterised by a non-trivial topology of the band structure, associated with the presence of Dirac cones at the corners of the Brillouin zone.
%,Wang2008,Wang2009,He2010,Khanikaev2013,Lu2014,Gao2015,

The existence of edge modes propagating in one preferential direction was firstly predicted and observed in photonic crystals \cite{RaghuHaldane2008}--\cite{Skirlo2015}, in analogy with electronic edge states in systems exhibiting the quantum Hall effect \cite{Klitzing1980}.
Uni-directional topologically protected  interfacial states have been identified in \cite{Siroki2017} for  photonic crystals composed of two regions, each possessing hexagonally arranged dielectric rods with different radii distributed on a triangular lattice.
The effect of singular points  on the boundaries of solids through which light propagates has been modelled in \cite{Luoetal} for the purpose of developing broadband energy harvesting techniques.
Analogues of topological insulators have also appeared in the study of optical lattices \cite{Goldmanetal}.
In plasmonics, the generation of surfaces modes  can be achieved by introducing structured inhomogeneities into the surface \cite{Pendryetal}. Periodic honeycomb plasmonics capable of supporting uni-directional edge states have appeared in \cite{Jinetal, Nalitov}.

%Yang2015,ChenWu2016,
In the context of classical mechanics, there are only a few examples where edge states in structured media have been observed.
Acoustic metamaterials supporting topologically protected sound modes have been designed by introducing circulating fluids into a lattice structure \cite{Ni2015}--\cite{Souslov2017} or by creating interfaces between two phononic crystals having different geometrical properties \cite{He2016}.
An array of acoustic resonators periodically distributed within a hexagonal lattice has been used in \cite{Khanikaevetal} to study  the robustness of topological edge wave propagation when various defects are embedded in the lattice.
The analysis of waves trapped along coastlines fitted with structured  barriers has been presented in
\cite{Evansetal}, while similar trapping phenomena have been analysed with an
asymptotic model for stratified fluids  in \cite{Adamouetal}.
Interfacial waves have been observed in  elastic metamaterials, consisting of  two slabs with  arrangements of defects at two different scales \cite{Mousavi2015}.
Topologically protected plates have been constructed  by attaching a hexagonal array of resonators possessing different masses  to the plate in order to break the inversion symmetry \cite{PalRuzzene2017}. Preferential directionality has been generated in lattice structures by modifying the tension of the springs connecting the particles \cite{KariyadoHatsugai2015} or by {locally changing} the arrangement of  masses at the junctions within the structure \cite{Vila2017}. Localised interfacial  modes for circular arrays of inclusions in membranes have been studied in \cite{Maling}.
The analogue of the quantum Hall effect for mechanical systems has been analytically and experimentally investigated in \cite{HS&Huber}, where topologically protected edges modes  have been realised for  finite lattice systems whose nodal points are connected to a system of coupled pendula.
This has been further explored in \cite{Huber}, where a review of  recent attempts in  bridging the gap  between quantum and classical mechanics for the purposes of designing topological metamaterials is presented.

In this paper, we design a system composed of fundamental mechanical elements capable of supporting and controlling {interfacial waveforms. In particular, we} consider a hexagonal array of masses connected to gyroscopic spinners at the junctions. Gyroscopic spinners are employed to break the time-reversal symmetry and alter the topology of the band-gaps in correspondence with the Dirac points. We will show how interfacial waves propagating in one preferential direction can be generated by dividing the domain of the lattice in two regions, where the spinners rotate in opposite directions. We will also demonstrate that the preferential direction can be inverted by not only reversing the direction of the spinners in separate regions, but also by changing the frequency of the harmonic {excitation} applied to a node along the interface.

The first model of an elastic gyro-system was proposed in \cite{Brun2012}, where both a monatomic and a biatomic lattice were analysed. The dispersion properties of a monatomic lattice with gyroscopic spinners were discussed in more detail in \cite{Carta2014}, where wave polarisation and standing waves were also investigated. It was demonstrated in \cite{Carta2017a} that in a gyro-lattice with two types of spinners, waves produced by a harmonic force with a specific frequency propagate along a single line, which can be diverted by changing the arrangement of the spinners within the medium. Furthermore, if waves are forced to travel along a closed path, their amplitude can be increased considerably ({an effect termed the ``DASER phenomenon''}).

The approach of \cite{Brun2012} was employed in \cite{Wang2015} to create topologically non-trivial edge waves, whose existence was demonstrated by numerical simulations in the transient regime. In this paper, attention is focussed instead on interfacial waveforms in a different lattice structure, for which it is possible to derive an analytical expression for the dispersion relation; moreover, the simulations are carried out in the time-harmonic regime to fully understand the frequency dependence of the phenomenon. Edge waves propagating in one direction in a gyroscopic metamaterial were observed experimentally in \cite{Nash2015}.

%D'Eleuterio1988,Yamanaka1996,HassanpourHeppler2016a,
Gyroscopic spinners confer a chiral nature to the system. ``Chirality'' is the property of an object that is not superimposable onto its mirror image \cite{Thomson1894}. Chirality has been employed in elastic lattices to create an effective auxetic medium \cite{PraLak1997,SpaRuz2012}, to alter the dispersive behaviour of a system \cite{Spadoni2009,BacigalupoGambarotta2016} and to generate negative refraction \cite{Zhu2014,Tallarico2016}. Chirality can be also implemented in continuous structural elements, including beams and plates, by using the concept of ``distributed gyricity''. In particular, the theory of gyro-elastic beams (or ``gyrobeams'') was proposed in \cite{D'EleuterioHughes1984} and developed in \cite{HughesD'Eleuterio1986}--\cite{HassanpourHeppler2016b}, where special attention was given to the study of the eigenfrequencies and eigenmodes of this structural element. In addition, in \cite{Carta2017b} it was shown that gyrobeams can be used {as an efficient tool} to reduce the vibrations of a structural system in the low-frequency regime.

The paper is organised as follows. In Section \ref{SectionHexagonalLattice}, we present the hexagonal chiral system, for which we derive the equations of motion (see Section \ref{SectionGoverningEquationsHexagonalLattice}) and the dispersion properties (see Section \ref{SectionDispersionHexagonalLattice}). We demonstrate how to generate interfacial waveforms in this medium, exploiting its topologically non-trivial band structure in Section \ref{SectionComsolHexagonalLattice}. Then, in Section \ref{SectionTriangularLattice}, we describe the more general case of a triangular gyro-lattice with links of different stiffness, of which the hexagonal chiral lattice in Section \ref{SectionHexagonalLattice} represents a degenerate case. After discussing the governing equations and the band diagrams of this system in Sections \ref{SectionGoverningEquationsTriangularLattice} and \ref{SectionDispersionTriangularLattice} respectively, we demonstrate that interfacial waveforms with preferential directionality can be also realised in this type of lattice, as detailed in Section \ref{SectionComsolTriangularLattice}. Finally, in Section \ref{Conclusions} we provide some concluding remarks.

\section{Hexagonal chiral lattice}
\label{SectionHexagonalLattice}

\begin{figure}%[!htcb]
\centering
\includegraphics[width=12cm]{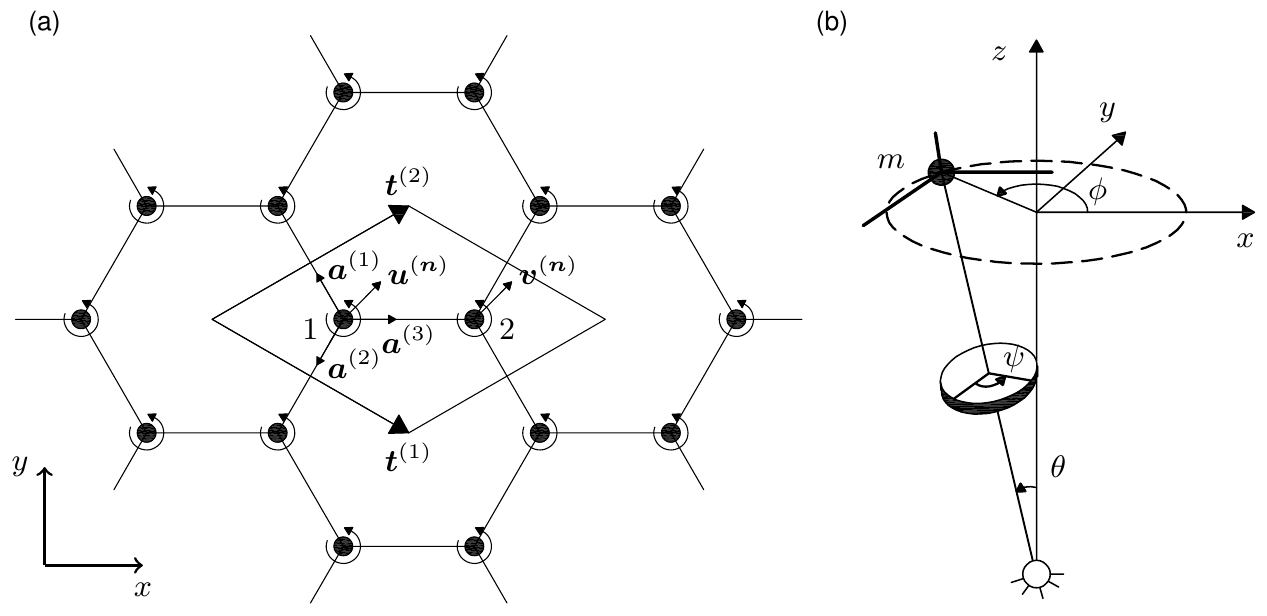}
\caption{\footnotesize (a) Hexagonal lattice connected to a uniform system of gyroscopic spinners, where ${\bm u}^{({\bm n})}$ and ${\bm v}^{({\bm n})}$ denote the displacements of the two lattice particles of the elementary cell; (b) representation of a gyroscopic spinner, where $\psi$, $\phi$ and $\theta$ are the angles of spin, precession and nutation respectively.}
\label{LatticeAndGyro}
\end{figure}

We consider a hexagonal lattice of masses $m$, connected by non-inertial elastic links of length $l$ and stiffness $c$, as shown in Fig. \ref{LatticeAndGyro}a. The masses of the lattice are attached to gyroscopic spinners, which are geometrically identical and spin with the same rate. Each  spinner is pinned at the bottom end, where it can rotate but it cannot translate. At the top end, it undergoes the same displacement as the lattice particle to which it is connected. A schematic representation of each gyroscopic spinner is illustrated in Fig. \ref{LatticeAndGyro}b, where the angles $\psi$, $\phi$ and $\theta$ are the angles of spin, precession and nutation respectively.

{In the undeformed configuration, the axis of each spinner is perpendicular to the $xy$-plane}.
When the lattice masses move due to an incoming wave, the spinners start precessing and exert a force that is perpendicular to the mass displacement. Assuming that $\theta \ll 1$ and gravity forces are negligible, the lattice particles are constrained to move in the $xy$-plane.

%%%%%%%%%%%%%%%%%%%%%%%%%%%%%%%%%%%%%%%%%%%%
%%%%%%%%%%%%%%%%%%%%%%%%%%%%%%%%%%%%%%%%%%%%

\subsection{Governing equations}
\label{SectionGoverningEquationsHexagonalLattice}

The vectors ${\bm t}^{(1)} = (3l/2, -\sqrt{3}l/2)^{\mathrm{T}}$ and ${\bm t}^{(2)} = (3l/2, \sqrt{3}l/2)^{\mathrm{T}}$, indicated in Fig. \ref{LatticeAndGyro}a, define the periodicity of the system. The position of each lattice particle is given by ${\bm x}^{({\bm n})} = {\bm x}^{({\bm 0})} + n_{x} {\bm t}^{(1)} + n_{y} {\bm t}^{(2)}$, where ${\bm n} = (n_{x},n_{y})^{\mathrm{T}}$ is the multi-index and ${\bm x}^{({\bm 0})}$ is the position of a reference particle in the lattice.

The elementary cell of the periodic structure consists of two particles, whose displacements are denoted by ${\bm u}^{({\bm n})}$ and ${\bm v}^{({\bm n})}$ (see Fig. \ref{LatticeAndGyro}a). The equations of motion of the two lattice particles in the time-harmonic regime are given by
\begin{subequations}\label{EquationsOfMotion}
\begin{equation}\label{EquationsOfMotiona}
-m \omega^{2} {\bm u}^{({\bm n})} = c \sum_{j = 1}^{3} \left[ {\bm a}^{(j)} \cdot \left( {\bm v}^{({\bm n}-{\bm e}_{j})} - {\bm u}^{({\bm n})} \right) \right] {\bm a}^{(j)} + \mathrm{i} \alpha \omega^{2} {\bm R} {\bm u}^{({\bm n})} \; ,
\end{equation}
\begin{equation}\label{EquationsOfMotionb}
-m \omega^{2} {\bm v}^{({\bm n})} = c \sum_{j = 1}^{3} \left[ {\bm a}^{(j)} \cdot \left( {\bm u}^{({\bm n}+{\bm e}_{j})} - {\bm v}^{({\bm n})} \right) \right] {\bm a}^{(j)} + \mathrm{i} \alpha \omega^{2} {\bm R} {\bm v}^{({\bm n})} \; ,
\end{equation}
\end{subequations}
where $\omega$ is the radian frequency, and the vectors ${\bm e}^{(1)} = (1,0)^{\mathrm{T}}$, ${\bm e}^{(2)} = (0,1)^{\mathrm{T}}$ and ${\bm e}^{(3)} = (0,0)^{\mathrm{T}}$ are used to specify the positions of the neighbouring particles. The unit vectors ${\bm a}^{(j)}$ in \eqref{EquationsOfMotion} define the directions of the lattice links (see Fig. \ref{LatticeAndGyro}a):
\begin{equation}\label{vectorsa}
{\bm a}^{(1)} = (-1/2, \sqrt{3}/2)^{\mathrm{T}} \; , \quad {\bm a}^{(2)} = (-1/2, -\sqrt{3}/2)^{\mathrm{T}} \; , \quad {\bm a}^{(3)} = (1, 0)^{\mathrm{T}} \; ,
\end{equation}
while the matrix ${\bm R}$ is the rotation matrix
\begin{equation}\label{matrixR}
\bm{R} =
\begin{pmatrix}
0 & 1 \\
-1 & 0 \\
\end{pmatrix}
\; .
\end{equation}
The parameter $\alpha$ in \eqref{EquationsOfMotion} represents the spinner constant, which was obtained in \cite{Brun2012} under the  assumption that the frequency of the nutation angle $\theta$ matches the radian frequency $\omega$ of the displacements of the lattice particles. The spinner constant $\alpha$ depends on the geometry and spin rate of the spinners \cite{Brun2012}.

The quasi-periodicity of the system is described by the Bloch-Floquet conditions:
\begin{equation}\label{BlochFloquetConditions}
{\bm W} \left({\bm r} + n_{x} {\bm t}^{(1)} + n_{y} {\bm t}^{(2)}\right) = {\bm W} \left({\bm r}\right) \mathrm{e}^{\mathrm{i} {\bm k} \cdot {\bm T} {\bm n}} \; ,
\end{equation}
where ${\bm W} = \left( u_{x}, u_{y}, v_{x}, v_{y} \right)^{\mathrm{T}}$ is the displacement vector, ${\bm r} = \left( x, y \right)^{\mathrm{T}}$ is the position vector, ${\bm k} = \left( k_{x}, k_{y} \right)^{\mathrm{T}}$ is the wavevector (or Bloch vector), and
\begin{equation}\label{matrixT}
\bm{T} = \left( {\bm t}^{(1)},{\bm t}^{(2)} \right) = l
\begin{pmatrix}
3/2 & 3/2 \\
-\sqrt{3}/2 & \sqrt{3}/2 \\
\end{pmatrix}
\; .
\end{equation}
By introducing \eqref{BlochFloquetConditions} into \eqref{EquationsOfMotion}, we obtain the following system of equations in matrix form:
\begin{equation}\label{EquationsOfMotionMatrixForm}
\left[ {\bm C} - \omega^{2} \left( {\bm M} - {\bm A} \right) \right] {\bm W} = {\bm 0} \; ,
\end{equation}
where ${\bm M} = m {\bm I}$ is the mass matrix (${\bm I}$ is the $4 \times 4$ identity matrix),
\begin{equation}\label{matrixA}
\bm{A} = \mathrm{i} \alpha
\begin{pmatrix}
0 & -1 & 0 & 0 \\
1 & 0 & 0 & 0 \\
0 & 0 & 0 & -1 \\
0 & 0 & 1 & 0 \\
\end{pmatrix}
\end{equation}
is the spinners matrix, and
\begin{equation}\label{matrixC}
\bm{C} = c
\begin{pmatrix}
\frac{3}{2} & 0 & -1 - \frac{\mathrm{e}^{-\mathrm{i} \eta} + \mathrm{e}^{-\mathrm{i} \gamma}}{4} & \frac{\sqrt{3}\left( \mathrm{e}^{-\mathrm{i} \eta} - \mathrm{e}^{-\mathrm{i} \gamma} \right)}{4} \\
0 & \frac{3}{2} & \frac{\sqrt{3}\left( \mathrm{e}^{-\mathrm{i} \eta} - \mathrm{e}^{-\mathrm{i} \gamma} \right)}{4} & -\frac{3\left( \mathrm{e}^{-\mathrm{i} \eta} + \mathrm{e}^{-\mathrm{i} \gamma} \right)}{4} \\
-1 - \frac{\mathrm{e}^{\mathrm{i} \eta} + \mathrm{e}^{\mathrm{i} \gamma}}{4} & \frac{\sqrt{3}\left( \mathrm{e}^{\mathrm{i} \eta} - \mathrm{e}^{\mathrm{i} \gamma} \right)}{4} & \frac{3}{2} & 0 \\
\frac{\sqrt{3}\left( \mathrm{e}^{\mathrm{i} \eta} - \mathrm{e}^{\mathrm{i} \gamma} \right)}{4} & -\frac{3\left( \mathrm{e}^{\mathrm{i} \eta} + \mathrm{e}^{\mathrm{i} \gamma} \right)}{4} & 0 & \frac{3}{2} \\
\end{pmatrix}
\end{equation}
is the stiffness matrix, where $\eta = \left( 3 k_{x} - \sqrt{3} k_{y} \right)l/2$ and $\gamma = \left( 3 k_{x} + \sqrt{3} k_{y} \right)l/2$. {We note that the determinant of ${\bm C}$ is zero for any value of the  Bloch vector. In the static problem, this results in a  non-unique solution.}

\subsection{Dispersion properties}
\label{SectionDispersionHexagonalLattice}

For non-trivial solutions of \eqref{EquationsOfMotionMatrixForm} to exist, the following equation must hold:
\begin{equation}\label{DispersionRelation}
\mathrm{det} \left[ {\bm C} - \omega^{2} \left( {\bm M} - {\bm A} \right) \right] = 0 \; ,
\end{equation}
which represents the dispersion relation of the chiral system. Eq. \eqref{DispersionRelation} is an algebraic equation of fourth order in $\omega^{2}$ and {embeds the action due the gyroscopic motion of the spinners through the matrix ${\bm A}$. Analogously, in \cite{HS&Huber} coupled pendula have been used to impose a gyroscopic action on a periodic medium,  resulting in topological insulation.}

We normalise \eqref{DispersionRelation} by introducing the non-dimensional scalar quantities $\tilde{\alpha} = \alpha / m$, $\tilde{\omega} = \omega \sqrt{m/c}$ and $\tilde{\bm k} = {\bm k} l$ and the non-dimensional matrices $\tilde{\bm C} = {\bm C}/c$ and $\tilde{\bm A} = {\bm A}/m$. This leads to the following normalised form of the dispersion relation:
\begin{equation}\label{DispersionRelationNormalised}
\mathrm{det} \left[ \tilde{\bm C} - \tilde{\omega}^{2} \left( {\bm I} - \tilde{\bm A} \right) \right] = 0 \; .
\end{equation}
In the following, the symbol ``tilde'' will be omitted for ease of notation.

One solution of \eqref{DispersionRelation} is $\omega = 0$ for any value of the wavevector ${\bm k}$. This solution represents a rigid-body motion of the system, which is {statically undetermined}. Rigid-body motions are prevented by introducing internal links, as shown in Section \ref{SectionTriangularLattice}; in this case, $\omega = 0$ is not a solution of the dispersion relation.

When $\alpha < 1$, \eqref{DispersionRelation} admits three real and positive solutions in $\omega$, which yield three dispersion surfaces. When $\alpha > 1$, two solutions for $\omega^{2}$ are negative and hence only one dispersion surface is defined.

\begin{figure}%[!htcb]
\centering
\includegraphics[width=12cm]{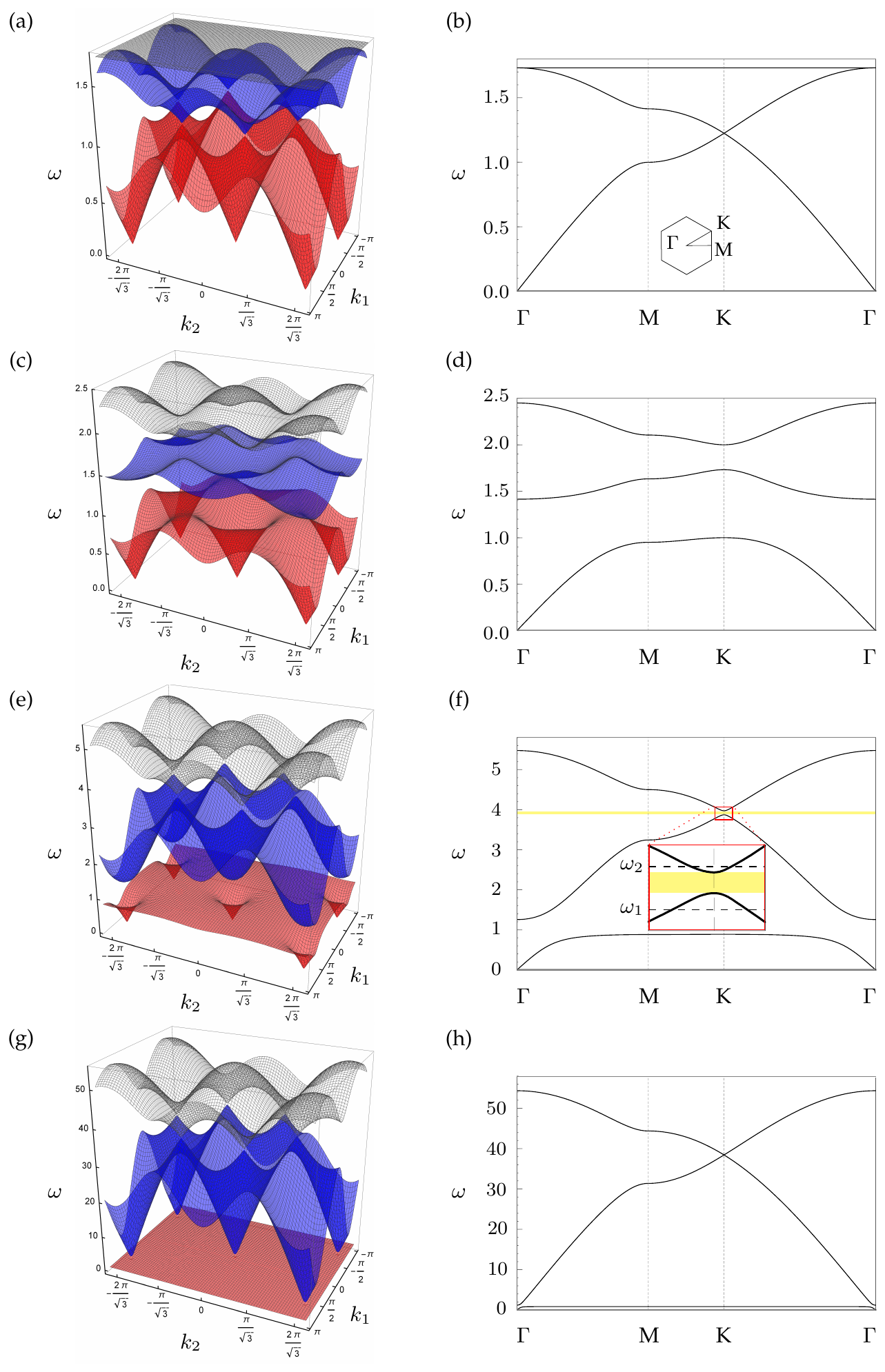}
\caption{\footnotesize Dispersion surfaces (a, c, e, g) and corresponding cross-sections (b, d, f, h), determined for $\alpha = 0$ (a, b), $\alpha = 0.5$ (c, d), $\alpha = 0.9$ (e, f), $\alpha = 0.999$ (g, h). For $\alpha <1$, the discrete hexagonal system exhibits three dispersion surfaces; the fourth solution of the dispersion relation, given by $\omega = 0$, is not shown in the figures for the sake of clarity. In (f) a magnified inset of the dispersion diagram is included. Note that the scales of the vertical axes are different, since the dispersion surfaces move to higher frequencies as $\alpha \to 1$.}
\label{DispersionSurfaces}
\end{figure}

Fig. \ref{DispersionSurfaces} shows the dispersion surfaces of the chiral hexagonal lattice for different values of the spinner constant $\alpha$. The diagrams on the right represent the cross-sections of the dispersion surfaces, calculated along the path $\Upgamma \mathrm{M K} \Upgamma$ in the reciprocal lattice, shown in the inset of Figs. \ref{DispersionSurfaces}b. The coordinates of the points $\Upgamma$, $\mathrm{M}$ and $\mathrm{K}$ are the following: $\Upgamma = (0,0)$, $\mathrm{M} = (2 \pi/(3 l),0)$, $\mathrm{K} = (2 \pi/(3 l),2 \pi/(3 \sqrt{3} l))$. The vectors ${\bm b}^{(1)}$ and ${\bm b}^{(2)}$, describing the periodicity of the reciprocal lattice, are given by
\begin{equation}\label{vectorsb}
\begin{array}{c}
{\bm b}^{(1)} = 2 \pi \frac{{\bm \tau}^{(2)} \times {\bm k}}{{\bm \tau}^{(1)} \cdot \left({\bm \tau}^{(2)} \times {\bm k}\right)} =
\begin{pmatrix}
2 \pi / 3 l \\
- 2 \pi / \sqrt{3} l \\ 0
\end{pmatrix}
\;,\\  \\
%\quad
{\bm b}^{(2)} = 2 \pi \frac{{\bm k} \times {\bm \tau}^{(1)}}{{\bm \tau}^{(1)} \cdot \left({\bm \tau}^{(2)} \times {\bm k}\right)} =
\begin{pmatrix}
2 \pi / 3 l \\
2 \pi / \sqrt{3} l \\ 0
\end{pmatrix}
\; ,
\end{array}
\end{equation}
where  ${\bm \tau}^{(j)}=(({\bm t}^{(j)})^{\rm T}, 0)^{\rm T}$, $j=1,2$, and ${\bm k}$ is the unit vector perpendicular to the lattice plane and directed along the positive $z$-axis.

%%%%%%%%%%%%%%%%%%%%%%%%%%%%%%%%%%%%%%%%%%%%
%%%%%%%%%%%%%%%%%%%%%%%%%%%%%%%%%%%%%%%%%%%%

When there are no spinners attached to the lattice ($\alpha = 0$), one solution is $\omega = \sqrt{3}$ for any value of the wavevector ${\bm k}$. The other two non-trivial solutions describe dispersion surfaces that intersect at a Dirac point, whose frequency is $\omega = \sqrt{3/2}$. We note that the dispersion diagrams in Figs. \ref{DispersionSurfaces}a and \ref{DispersionSurfaces}b are identical to those determined in \cite{CsertiTichy2004}.

When $0 < \alpha < 1$, the lowest  {nonzero} dispersion surface decreases as $\alpha$ is increased, while the highest two dispersion surfaces increase. As a consequence, two finite stop-bands appear within the band diagram of the gyro-lattice. Furthermore, the highest dispersion surface is not constant {in the $k_1 k_2$-plane}. The width of the upper  finite stop-band tends to zero as $\alpha \to 1^{-}$, and a new Dirac point originates. {The frequency of the new Dirac point is larger than that encountered for $\alpha =0$}. Figs. \ref{DispersionSurfaces}g and \ref{DispersionSurfaces}h show the dispersion diagrams for $\alpha = 0.999$; in this case,
{it is apparent that a new Dirac point is forming}, and  that the lowest non-zero dispersion surface is almost flat, except in a small proximity of point $\Upgamma$.

\begin{figure}%[!htcb]
\centering
\includegraphics[width=12cm]{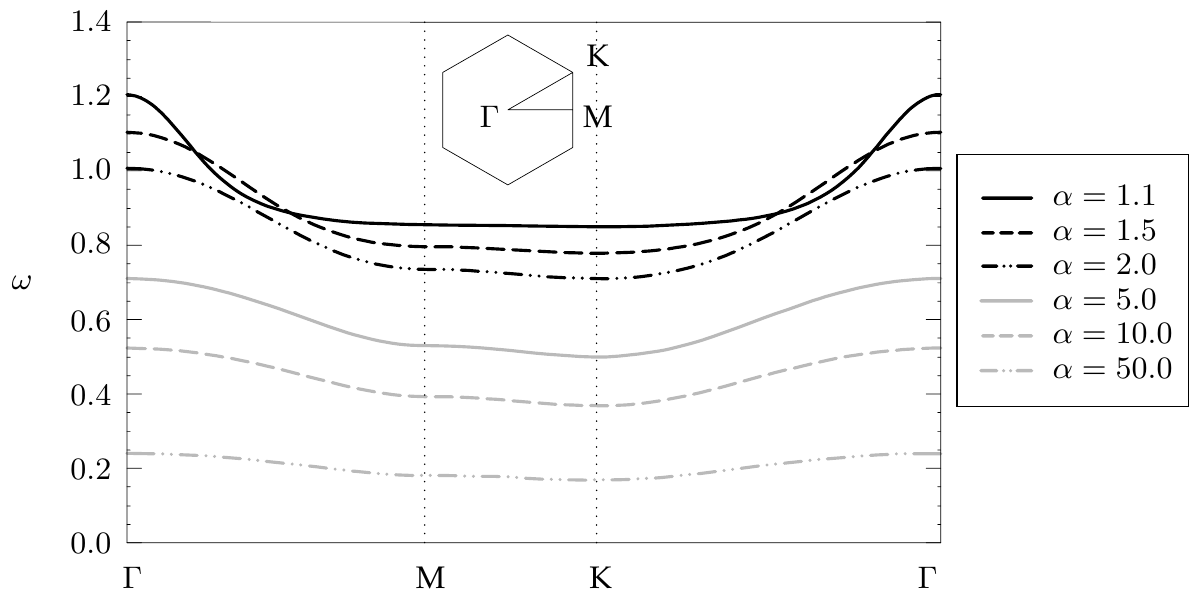}
\caption{\footnotesize Cross-sections of the dispersion surfaces of the hexagonal chiral lattice, computed for different values of $\alpha > 1$. The discrete hexagonal system is characterised by only one dispersion surface when $\alpha > 1$; another solution of the dispersion relation is $\omega = 0$ for any value of the wavevector ${\bm k}$.}
\label{DispersionSurfaces2}
\end{figure}

When $\alpha > 1$ there is only one dispersion surface, as discussed above. A similar phenomenon was observed for the triangular lattice with gyroscopic spinners \cite{Brun2012,Carta2014}, where there is only one  dispersion surface in the supercritical regime $\alpha > 1$. In contrast to  the triangular lattice, the dispersion surface of the hexagonal lattice, for $\alpha > 1$, does {not} pass through the origin because the hexagonal structure does not support shear waves. Fig. \ref{DispersionSurfaces2} shows that the dispersion surface decreases and becomes flatter as the spinner constant increases; in the limit when $\alpha \to \infty$, it tends to zero for any value of the wavevector ${\bm k}$.

%%%%%%%%%%%%%%%%%%%%%%%%%%%%%%%%%%%%%%%%%%%%
%%%%%%%%%%%%%%%%%%%%%%%%%%%%%%%%%%%%%%%%%%%%

\subsection{Waveforms with preferential directionality}
\label{SectionComsolHexagonalLattice}

In this section, we show the results of numerical simulations, where a harmonic displacement is applied to a particle in the hexagonal chiral lattice. The imposed displacement is ${\bm U}_{0} \mathrm{e}^{-\mathrm{i}\omega t}$, where $|{\bm U}_{0}| = 1$ is the amplitude and $t$ is time. The gyroscopic spinners are used to break the time-reversal symmetry and hence waves are expected to propagate along a preferential direction.

The time-harmonic response of the chiral lattice is determined using
\emph{Comsol Multiphysics} (version 5.2a). The gyroscopic effect is taken into account by introducing a force proportional to the particle displacement at each lattice junction (see the last terms of Eqs. \eqref{EquationsOfMotion}). In order to prevent rigid-body motions {of the whole system}, the displacements of the masses at the corners of the model are set equal to zero. In addition, PML (\emph{Perfectly Matched Layers}) are attached to the sides of the model to avoid reflections of waves generated by the harmonic displacement. In this way, the system is modelled as a spatially infinite medium. PML are created by using viscous dampers, whose parameters are tuned to minimise the reflection coefficient \cite{Carta2013,Carta2014}.

\subsubsection{Equations of motion of the hexagonal lattice connected to a heterogeneous system of spinners and subjected to an imposed displacement}

We consider the time-harmonic form of the governing equations  for a hexagonal lattice, connected to a non-uniform distribution of spinners, and we apply a harmonic displacement to {a lattice particle}.

We assume the points in the inhomogeneous lattice are grouped as pairs in cells defined by $t^{(1)}$ and $t^{(2)}$ of Figure \ref{LatticeAndGyro}(a). Within  the cell defined by the multi-index  ${\bm n^*}=(n_1^*, n_2^*)^{\rm T}$, we apply the condition
\[\delta_{j1} {\bm u}^{({\bm n})}+\delta_{j2} {\bm v}^{({\bm n})}= {\bm U}_0\;, \quad |{\bm U}_0|=1\;, \quad {\bm n}={\bm n^*}\;, \]
where $\delta_{ij}$ is the Kronecker delta  and $j=1$ or $2$ will correspond to a harmonic displacement  applied to the node labelled 1 or 2, respectively, in the cell of Figure \ref{LatticeAndGyro}(a).

Next we write the equations of motion for  the nodes in the ambient lattice, i.e. for all  nodes except for the $j$th node inside the cell  with  ${\bm n}= {\bm n}^*$.
We introduce the vector 
\[{\bm Y}_{\bm n}=( ({\bm u}^{({\bm n}+{\bm e}_1)})^{\rm T}, ({\bm u}^{({\bm n}+{\bm e}_2)})^{\rm T}, ({\bm u}^{({\bm n})})^{\rm T}, ({\bm v}^{({\bm n})})^{\rm T}, ({\bm v}^{({\bm n}-{\bm e}_1)})^{\rm T}, ({\bm v}^{({\bm n}-{\bm e}_2)})^{\rm T})^{\rm T}\] and then the time-harmonic equations of motion for the nodes in the elementary cell corresponding to the multi-index ${\bm n}=(n_1, n_2)^{\rm T}$ are written as
\[{\bm K}_{\bm n}{\bm Y}_{\bm n}={\bm 0}\;,\]
where
%${\bm F}_j=[(F_1, F_2)^{\rm T}\delta_{i, j}]_{i=1}^2$, $j=1,2$, and
${\bm K}_{\bm n}$ is a $4 \times 12$ block matrix taking the form
\[{\bm K}_{\bm n}=[
{\bm A}^{(1)}, {\bm A}_{\bm n}, {\bm A}^{(2)}
]\;,\]
with
\[{\bm A}^{(1)}=-c\left[\begin{array}{cc}{\bm 0}_2 & {\bm 0}_2\\
{\bm a}^{(1)} \otimes {\bm a}^{(1)} & {\bm a}^{(2)} \otimes {\bm a}^{(2)}
\end{array} \right]\;, \quad {\bm A}^{(2)}=\left[\begin{array}{cc}{\bm 0}_2 & {\bm I}_2 \\
 {\bm 0}_2 &{\bm 0}_2
\end{array}\right]{\bm A}^{(1)} \]
and
\[{\bm A}_{\bm n}=\left[\begin{array}{cc}{\bm T}_1({\bm n}) & -c\,{\bm a}^{(3)} \otimes {\bm a}^{(3)} \\
 -c\,{\bm a}^{(3)} \otimes {\bm a}^{(3)} &{\bm T}_2({\bm n})
\end{array} \right]\;,\]
{where}
\[{\bm T}_j=- \omega^2( m {\bm I}_2 +{\rm i}\alpha_j({\bm n})  {\bm R})+c\sum_{i=1}^3  {\bm a}^{(i)} \otimes  {\bm a}^{(i)} \;, \quad j=1, 2\;.\]
Here, ${\bm I}_n$ and ${\bm 0}_n$   are the $n\times n$ identity  and null matrices, respectively.
As we consider a non-uniform distribution of spinners in the simulations, in the above formulation we have introduced the  $\alpha_j({\bm n})$ to denote the spinner constant for the spinner attached to the $j$th node in the elementary cell associated with the multi-index ${\bm n}$ (see Fig. \ref{LatticeAndGyro}). We allow for the possibility that such constants may change from node to node in the lattice.

\subsubsection{Uni-directional localised waveforms for different types of interfaces}

Using a variety of illustrations, we show that localised interfacial waveforms can be supported in an inhomogeneous hexagonal lattice attached to gyroscopic spinners.
In particular, the inhomogeneity of the lattice is brought by the spatial dependence of the spinner constants, which are always chosen equal to $\pm 0.9$.
In addition, we demonstrate that the direction of these waveforms can be controlled by altering the spin directions of the gyroscopic spinners or changing the frequency of the applied harmonic displacement.

In the first class of simulations, we divide the lattice domain into two different regions. The regions are distinguished by having gyroscopic spinners characterised by spinner constants of the same magnitude but of opposite {sign}. At specific frequencies, which can be predicted from the dispersion analysis described in Section \ref{SectionDispersionHexagonalLattice}, interfacial waves propagate along the internal boundaries between the two different regions.
In these computations, the lattice is a $150 \, l \times 110 \, l$ rectangle.

%%%%%%%%%%%%%%%%%%%%%%%%%%%%%%%%%%%%%%%%%%%%
\begin{figure}%[!htcb]
\centering
\includegraphics[width=12cm]{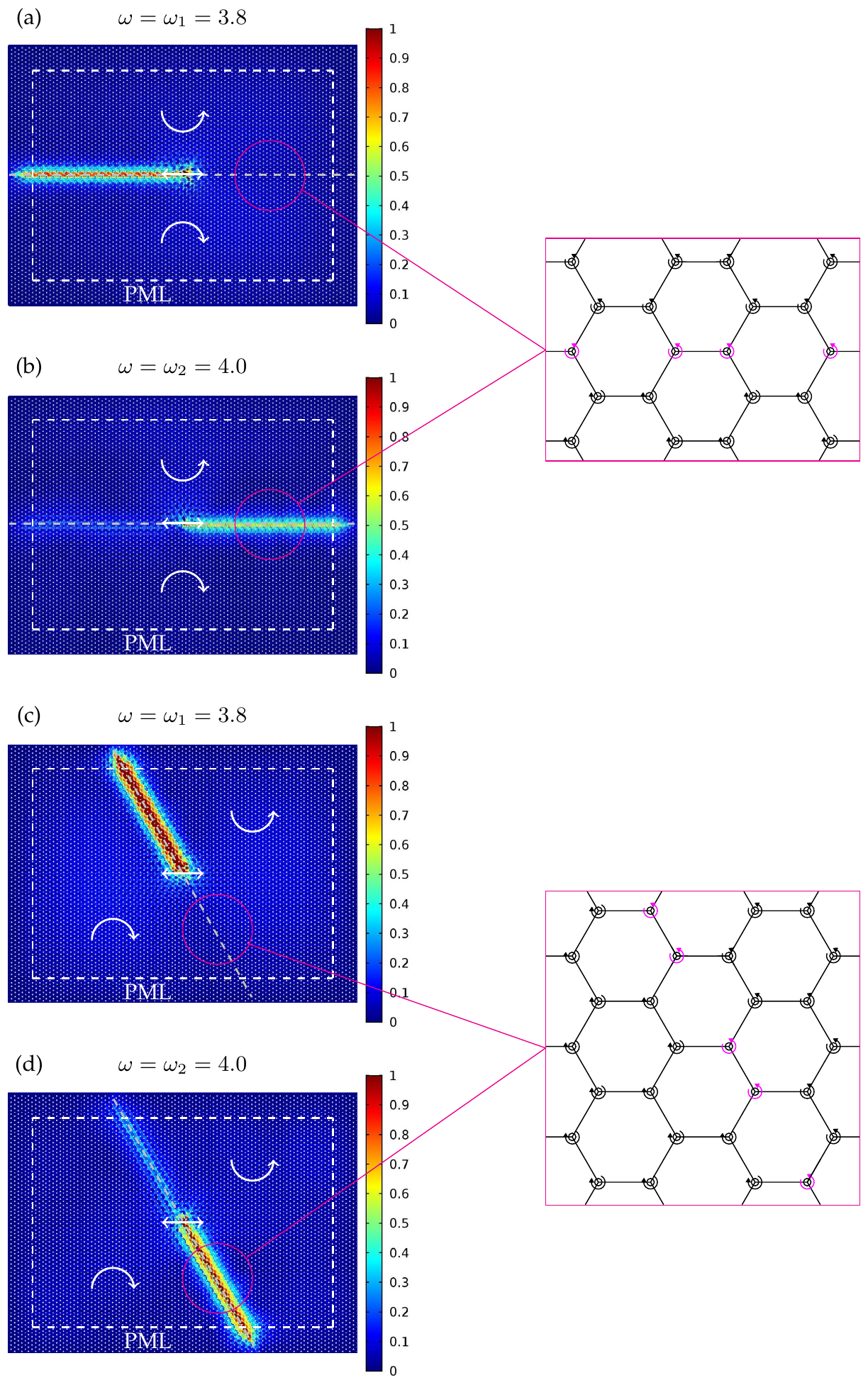}
\caption{\footnotesize Displacement amplitude fields in the chiral lattice at different frequencies of the applied harmonic displacement and for different spin directions of the spinners (indicated by the white circular arrows). The harmonic displacement (represented by a white arrow) is applied to the central node of the lattice. In these computations, the absolute value of the spinner constant is $\left|\alpha\right| = 0.9$, and the frequency of the harmonic displacements is (a,c) $\omega = \omega_1 = 3.8$ and (b,d) $\omega = \omega_2 = 4.0$ (see also Fig. \ref{DispersionSurfaces}f). PML (\emph{Perfectly Matched Layers}) are attached to the sides of the model, as indicated by the dashed white lines. The interfaces between regions where spinners rotate in opposite directions are represented by dashed grey lines. To the right of each computation we provide a zoom of the interface.}
\label{InterfacialWaves1}
\end{figure}
%%%%%%%%%%%%%%%%%%%%%%%%%%%%%%%%%%%%%%%%%%%%

In Fig. \ref{InterfacialWaves1} we present several computations for a hexagonal lattice composed of two regions attached to different sets of spinners. Each computation is accompanied by a close up of how the interface between the two regions is formed.
We note that there are many ways to design the interface of these regions{, but} we have chosen for our illustrations interfaces involving nodes connected to spinners.
In {Figs. \ref{InterfacialWaves1}a {and} \ref{InterfacialWaves1}b}, we divide the lattice domain into two regions separated by a horizontal line. In the upper and lower regions, gyroscopic spinners have the same spin rate ($\left|\alpha\right| = 0.9$) but they spin in opposite directions, as indicated by the white circular arrows. At the central node of the domain we impose a harmonic displacement, whose direction is represented by a white arrow.  Fig. \ref{InterfacialWaves1}a illustrates the displacement field calculated at the radian frequency $\omega = \omega_1 = 3.8$ for the applied displacement, which is below the lower limit of the stop-band highlighted in yellow in Fig. \ref{DispersionSurfaces}f, related to the problem concerning free vibrations in the homogeneous system. From Fig. \ref{InterfacialWaves1}a, it is apparent that waves propagate along the interface between the two regions in one direction. We mention that this direction  can be {reversed} by swapping the spin directions of the spinners. Fig. \ref{InterfacialWaves1}b shows the interfacial waves generated at the  radian frequency $\omega = \omega_2 = 4.0$ for the applied displacement. This frequency is above the upper limit of the stop-band coloured in yellow in Fig. \ref{DispersionSurfaces}f. Comparing Figs. \ref{InterfacialWaves1}a {and} \ref{InterfacialWaves1}b, we observe that when the frequency of the harmonic displacement is chosen in the proximity of the upper limit of the stop-band in \ref{DispersionSurfaces}f, waves propagate in the opposite direction with respect to the case when the frequency is close to the lower limit of this stop-band.

Figs. \ref{InterfacialWaves1}c {and} \ref{InterfacialWaves1}d show the interfacial waves in a lattice, where the two regions containing spinners that spin in opposite directions are separated by a line inclined by $-\pi/3$ with respect to the $x$-axis. The direction of the force is the same as in Figs. \ref{InterfacialWaves1}a and \ref{InterfacialWaves1}b. Again, these computations illustrate that the direction of propagation can be changed by modifying the frequency of the harmonic displacement. In addition, they show that the phenomenon is independent of the direction of the external excitation.

It is interesting to note that interfacial waves propagating in the vertical direction cannot be produced in such a structure, because there are no links connecting two adjacent masses that are oriented along the $y$-axis.
In all the examples of Figure \ref{InterfacialWaves1}, the waveforms appear to be exponentially  localised and the strength of this localisation appears to be uniform along the interface. We also observe that if the spin directions of the spinners on the interface are reversed, the wave pattern is not affected significantly. Furthermore, we note that waves do not propagate if the nodes at the interface are not connected to spinners, since the frequencies chosen for the numerical simulations lie within the stop-band of the hexagonal lattice without spinners.

%%%%%%%%%%%%%%%%%%%%%%%%%%%%%%%%%%%%%%%%%%%%
\begin{figure}%[!htcb]
\centering
\includegraphics[width=12cm]{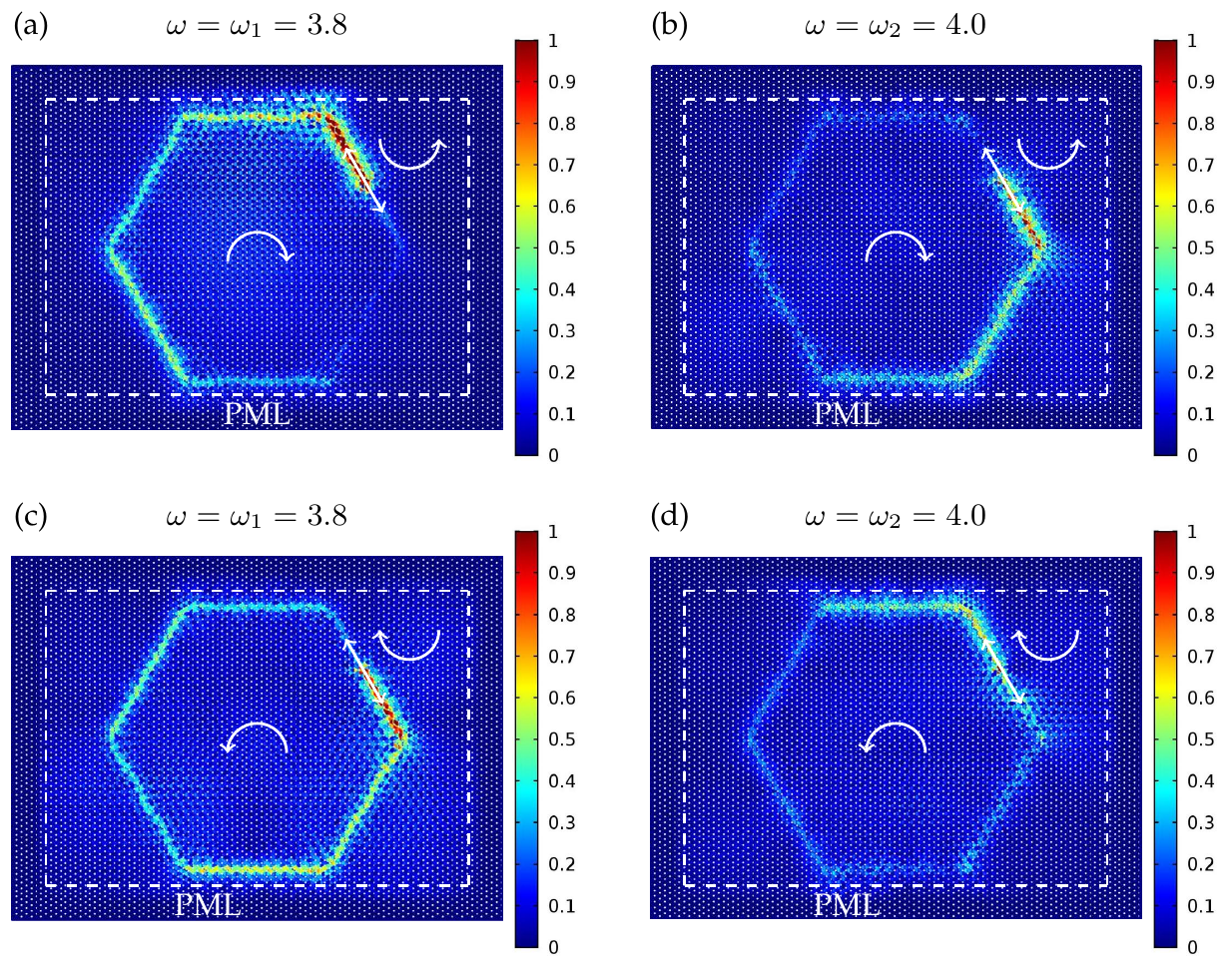}
\caption{\footnotesize Interfacial waveforms with preferential directionality in a hexagonal lattice connected to a system of gyroscopic spinners. The lattice domain is divided into two regions: in the hexagonal region the spinners rotate (a,b) clockwise and (c,d) anticlockwise, while the ambient medium possesses spinners which rotate in the opposite direction to those situated inside the hexagonal region. The absolute value of the spinner constant is $\left|\alpha\right| = 0.9$ for all the gyroscopic spinners. A harmonic displacement, represented by the straight arrow, is applied to a node on the interface of these regions.
The displacement amplitude fields are obtained for an applied displacement with a frequency (a,c) $\omega = \omega_1 = 3.8$ and (b,d) $\omega = \omega_2 = 4.0$ (see also inset in Fig. \ref{DispersionSurfaces}f).
Localised waveforms with preferential directionality are observed, whose intensities vary along the hexagonal boundary. The figures illustrate that the preferential direction of the interfacial waveform can be influenced by interchanging the direction of rotation of the spinners or by changing the frequency of the external excitation.}
\label{InterfacialWaves6}
\end{figure}
%%%%%%%%%%%%%%%%%%%%%%%%%%%%%%%%%%%%%%%%%%%%

Another example of an interfacial waveform is presented in Fig. \ref{InterfacialWaves6}. In this case, the interface has a hexagonal shape. Figs. \ref{InterfacialWaves6}a and \ref{InterfacialWaves6}b show the displacement fields computed for two different values of the frequency of the imposed harmonic displacement, specified at the top of each figure (see also Fig. \ref{DispersionSurfaces}f); in both Figs. \ref{InterfacialWaves6}a and \ref{InterfacialWaves6}b, the spinners inside the hexagonal domain rotate clockwise, while the spinners in the ambient medium rotate anticlockwise. The comparison between Figs. \ref{InterfacialWaves6}a and \ref{InterfacialWaves6}b emphasises the dependance of the preferential directionality of the medium on the  frequency of the harmonic displacement, which can be predicted from the dispersion analysis discussed in Section \ref{SectionDispersionHexagonalLattice}.

%In order to better visualise this phenomenon, we illustrate the motion of the particles in proximity of the harmonic displacement with two videos, included in the Supplementary Material. The videos show a narrow region of the gyro-elastic lattice, where the dashed red line represents the interface and the blue arrow indicates the direction of the applied harmonic displacement. It is clear that in the video corresponding to the case presented in Fig. \ref{InterfacialWaves6}a, the displacements of the lattice particles are significant in the direction along the red line above the applied harmonic displacement, while they are small below the point where this displacement is applied. The opposite scenario is observed in the video related to the configuration in Fig. \ref{InterfacialWaves6}b.

In Figs. \ref{InterfacialWaves6}c and \ref{InterfacialWaves6}d, the spinners inside the hexagonal lattice spin anticlockwise, while those in the ambient medium spin clockwise. Comparing Figs. \ref{InterfacialWaves6}c and \ref{InterfacialWaves6}a and Figs. \ref{InterfacialWaves6}d and \ref{InterfacialWaves6}b, it is apparent that interchanging the directions of spin for the gyroscopes reverses the direction of the interfacial waveform. In addition, we note that in all the simulations of Fig. \ref{InterfacialWaves6} the intensity of the waveform varies moving along the boundary between the two regions, due to the scattering occurring at the points of slope discontinuity of the interface.

%%%%%%%%%%%%%%%%%%%%%%%%%%%%%%%%%%%%%%%%%%%%
\begin{figure}%[!htcb]
\centering
\includegraphics[width=12cm]{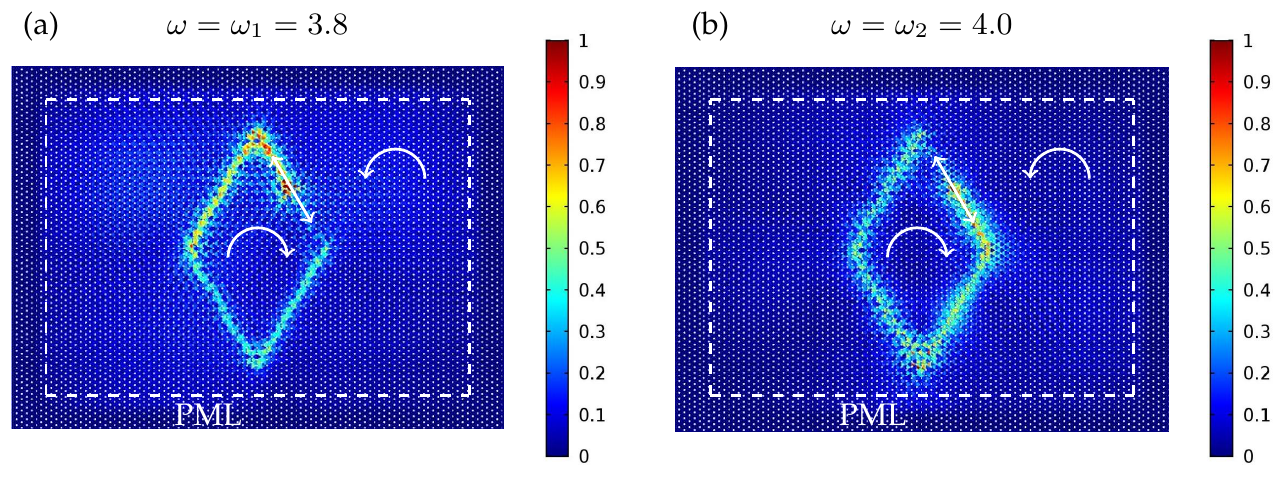}
\caption{Another example showing the sensitivity of the preferential directionality of an interfacial waveform in a hexagonal lattice on spinners, containing a subdomain where the spinners rotate in opposite direction to those in the ambient medium. The radian frequency of the applied harmonic displacement is (a) $\omega = \omega_1 = 3.8$ and (b) $\omega = \omega_2 = 4.0$ (see also Fig. \ref{DispersionSurfaces}f). As in the previous simulations, the absolute value of the spinner constant is $\left|\alpha\right| = 0.9$ throughout the domain.}
\label{InterfacialWaves2}
\end{figure}
%%%%%%%%%%%%%%%%%%%%%%%%%%%%%%%%%%%%%%%%%%%%

The effect shown in Fig. \ref{InterfacialWaves6} is not unique to the hexagonal subdomain within the inhomogeneous lattice.
A final illustration of an interfacial waveform along the boundary of a different subset of an inhomogeneous hexagonal lattice is given in Fig. \ref{InterfacialWaves2}, where the internal boundary is now a rhombus.
In particular, Figs. \ref{InterfacialWaves2}a and \ref{InterfacialWaves2}b are obtained for a radian frequency of the applied displacement equal to $\omega = 3.8$ and $\omega = 4.0$, respectively. These frequencies are indicated by $\omega_1$ and $\omega_2$, respectively, in Fig. \ref{DispersionSurfaces}f.
Once again, in changing the frequency of the external excitation, it is possible to alter the direction of the waveform.
Moreover, we note that as the rhombus chosen has fewer vertices and a smaller area than the hexagon used in the computations of Fig. \ref{InterfacialWaves6}, we do not observe a significant drop in the intensity of the waveform as in the examples presented there.

\section{A degenerating triangular chiral lattice}
\label{SectionTriangularLattice}

The hexagonal lattice described in Section \ref{SectionHexagonalLattice} is statically undetermined and its dispersion diagram is characterised by a single acoustic branch. In order to allow for propagation of both shear and pressure waves in the medium, we introduce internal links in the structure, which then becomes a triangular lattice. This lattice degenerates into the hexagonal lattice of Section \ref{SectionHexagonalLattice} when the stiffness of the internal links tends to zero.

\begin{figure}%[!htcb]
\centering
\includegraphics[width=12cm]{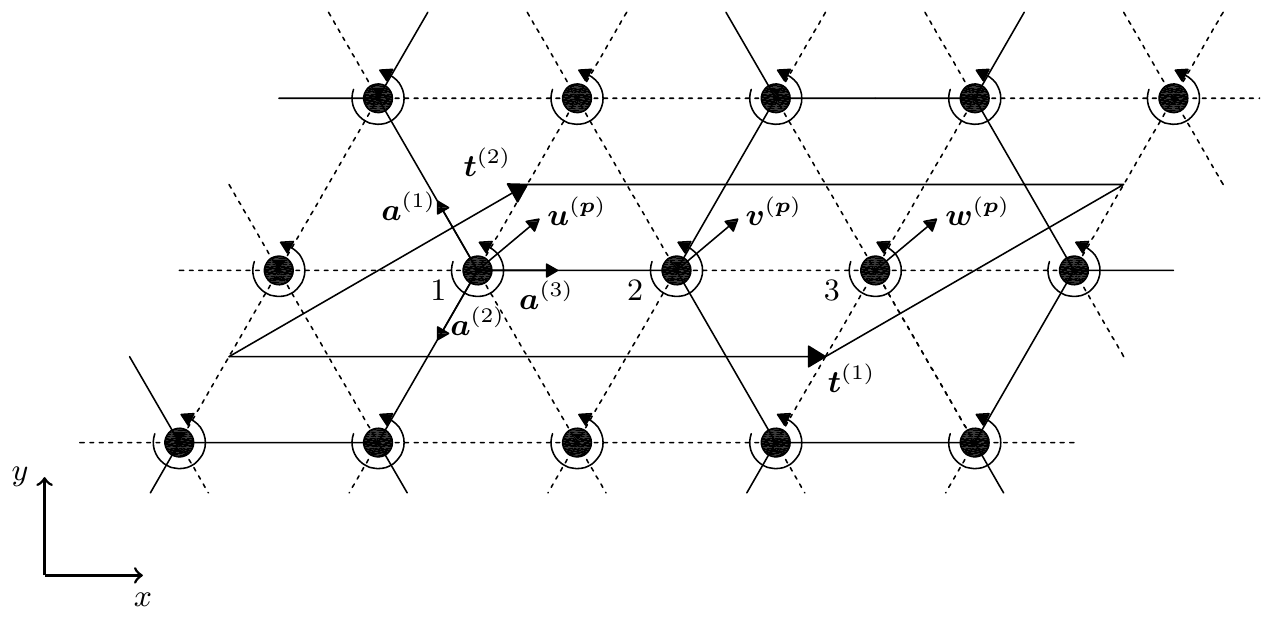}
\caption{\footnotesize An inhomogeneous triangular lattice structure linked to a system of spinners. The elementary cell for this structure is shown, which contains 3 particles having displacements  ${\bm u}^{({\bm p})}$, ${\bm v}^{({\bm p})}$ and ${\bm w}^{({\bm p})}$. The lattice is composed of springs with stiffnesses $c$ (solid lines) and $c_\varepsilon$ (dashed lines), where $c_\varepsilon \ll c$. Springs possessing  stiffness $c_\epsilon$ form the internal connections of hexagons in the lattice. Here the lattice basis vectors are taken as ${\bm t}^{(1)} = (3l,0)^{\mathrm{T}}$ and ${\bm t}^{(2)} = ( 3 l/2, \sqrt{3}l/2)^{\mathrm{T}}$.}
\label{LatticeAndGyro2}
\end{figure}

The triangular lattice is shown in Fig. \ref{LatticeAndGyro2}. It is composed of particles with mass $m$ and two types of springs, each possessing length $l$ and having stiffnesses $c$ and $c_\varepsilon$. The stiffness $c_\varepsilon$ is assumed to be small in comparison to $c$. As before, each mass is connected to a spinner that has a spinner constant $\alpha$.

%%%%%%%%%%%%%%%%%%%%%%%%%%%%%%%%%%%%%
%%%%%%%%%%%%%%%%%%%%%%%%%%%%%%%%%%%%%%%%%%%%%%

The elementary cell for this configuration is shown in Fig. \ref{LatticeAndGyro2}, where it can be seen this cell contains three particles, whose displacements {are denoted as} ${\bm u}^{({\bm p})}$, ${\bm v}^{({\bm p})}$ and ${\bm w}^{({\bm p})}$. In addition, the basis vectors ${\bm t}^{(1)} = (3l,0)^{\mathrm{T}}$ and ${\bm t}^{(2)} = ({3}l/2, \sqrt{3}l/2)^{\mathrm{T}}$ are used to define the periodicity of the system.
The unit vectors ${\bm a}^{(j)}$ introduced in (\ref{vectorsa}) will also be utilised to define the directions of the links in the lattice. A particle's position in this lattice can be determined through ${\bm x}^{({\bm p})} = {\bm x}^{({\bm 0})} + p_{x} {\bm t}^{(1)} + p_{y} {\bm t}^{(2)}$, where ${\bm p}=(p_x, p_y)^{\rm T}$ is a  multi-index.

\subsection{Governing equations of the heterogeneous triangular lattice}
\label{SectionGoverningEquationsTriangularLattice}

In the time-harmonic regime, the governing equations of the three particles in the elementary cell are
\begin{subequations}\label{EquationsOfMotionDTL}
\begin{eqnarray}
-m \omega^{2} {\bm u}^{({\bm p})}&=&
 \sum_{j=1}^3 \Big[  {\bm a}^{(j)} \cdot \Big(c({\bm v}^{({\bm p}-{\bm q}_{j})}- {\bm u}^{({\bm p})})\nonumber \\&&+c_\varepsilon({\bm w}^{({\bm p}+{\bm q}_{j}-{\bm e}_1)}- {\bm u}^{({\bm p})} )\Big) \Big] {\bm a}^{(j)} + \mathrm{i} \alpha \omega^{2} {\bm R} {\bm u}^{({\bm p})}\;, \nonumber \\ \label{EquationsOfMotionDTLa}
\end{eqnarray}
\begin{eqnarray}
-m \omega^{2} {\bm v}^{({\bm p})}&=&
\sum_{j=1}^3 \Big[  {\bm a}^{(j)} \cdot \Big(c({\bm u}^{({\bm p}+{\bm q}_{j})}- {\bm v}^{({\bm p})})\nonumber \\ &&+c_\varepsilon ({\bm w}^{({\bm p}-{\bm q}_{j})}- {\bm v}^{({\bm p})})\Big) \Big] {\bm a}^{(j)} + \mathrm{i} \alpha \omega^{2} {\bm R} {\bm v}^{({\bm p})}\;, \nonumber \\ \label{EquationsOfMotionDTLb}
\end{eqnarray}
\begin{eqnarray}
&&-m \omega^{2} {\bm w}^{({\bm p})}\nonumber \\&=& c_\varepsilon \sum_{j=1}^3 \left[ {\bm a}^{(j)} \cdot \left({\bm v}^{({\bm p}+{\bm q}_j)}+{\bm u}^{({\bm p}-{\bm q}_j+{\bm e_1})} - 2{\bm w}^{({\bm p})} \right) \right] {\bm a}^{(j)}+ \mathrm{i} \alpha \omega^{2} {\bm R} {\bm w}^{({\bm p})} \; , \nonumber \\ \label{EquationsOfMotionDTLc}
\end{eqnarray}
\end{subequations}
where ${\bm q}_1= {\bm e}_{1}-{\bm e}_{2}$, ${\bm q}_2={\bm e}_{2}$, ${\bm q}_3= 0{\bm e}_{1}+0{\bm e}_{2}$
and  ${\bm R}$ is the rotation matrix in (\ref{matrixR}). We note that $(\ref{EquationsOfMotionDTLa})$ and $(\ref{EquationsOfMotionDTLb})$ coincide with (\ref{EquationsOfMotiona}) and (\ref{EquationsOfMotionb}) if $c_\varepsilon=0$, while in this case $(\ref{EquationsOfMotionDTLc})$ implies {${\bm w}^{(\bm p)}$ }vanishes for $\omega > 0$.

We  proceed to analyse the Bloch-Floquet modes of the above system by introducing the quasi-periodicity conditions described in (\ref{BlochFloquetConditions})--(\ref{matrixT}), where ${\bm W}= \left( u_{x}, u_{y}, v_{x}, v_{y}, w_{x}, w_{y} \right)^{\mathrm{T}}$ is the new displacement vector, ${\bm T} {= \left( {\bm t}^{(1)},{\bm t}^{(2)} \right)}$ is constructed from the new lattice basis vectors and ${\bm p}$ replaces ${\bm n}$. In a similar way to that in Section \ref{SectionHexagonalLattice}, we then arrive at
a system of equations for ${\bm W}$ in the form
\begin{equation}\label{EquationsOfMotionMatrixForm1}
\left[ {\bm C}_\varepsilon - \omega^{2} \left( {\bm M} - {\bm A} \right) \right] {\bm W} = {\bm 0} \; ,
\end{equation}
where ${\bm M}=m{\bm I_6}$ (${\bm I_j}$ is the $j\times j$ identity matrix), the spinners matrix
\[{\bm A}={\rm i} \alpha\,  \text{diag}\left( \left(\begin{array}{ll}0 &-1\\
1 &\>\>\>0
\end{array}\right), \left(\begin{array}{ll}0 &-1\\
1 &\>\>\>0
\end{array}\right), \left(\begin{array}{ll}0 &-1\\
1 &\>\>\>0
\end{array}\right)\right)\;,\]
and  ${\bm C}_\varepsilon$ is a $6 \times 6$ stiffness matrix depending on the  wavevector ${\bm k}$ given by
\begin{equation}\label{matrixC_e}
\bm{C}_\varepsilon =\begin{pmatrix}
\bm{C}_\varepsilon^{(1)} & \bm{C}_\varepsilon^{(3)}\\
\overline{\bm{C}_\varepsilon^{(3){\rm T} }}& \bm{C}_\varepsilon^{(2)}\\
\end{pmatrix},
\end{equation}
with
\begin{equation}\label{matrixC_e^1}
{\bm C}_\varepsilon^{(1)}={\bm C}+\frac{3}{2} c \varepsilon {\bm I}_4\;, \quad {\bm C}_\varepsilon^{(2)}= 3 c \varepsilon {\bm I}_2\;,
\end{equation}
and
\begin{equation}\label{matrixC_e^2}
{\bm C}_\varepsilon^{(3)} =\frac{c \varepsilon}{4} \begin{pmatrix}
-(4\mathrm{e}^{-\mathrm{i} \mu}+\mathrm{e}^{-\mathrm{i} \eta} + \mathrm{e}^{-\mathrm{i} \gamma}) & \sqrt{3}\left(-\mathrm{e}^{-\mathrm{i} \eta} + \mathrm{e}^{-\mathrm{i} \gamma} \right)\\
-\sqrt{3}\left( \mathrm{e}^{-\mathrm{i} \eta} - \mathrm{e}^{-\mathrm{i} \gamma} \right)& -3\left( \mathrm{e}^{-\mathrm{i} \eta} +\mathrm{e}^{-\mathrm{i} \gamma} \right)\\
-(4 +\mathrm{e}^{-\mathrm{i} \eta} + \mathrm{e}^{-\mathrm{i}\gamma})&  \sqrt{3}\left( \mathrm{e}^{-\mathrm{i} \eta} - \mathrm{e}^{-\mathrm{i} \gamma} \right)\\
\sqrt{3} \left( \mathrm{e}^{-\mathrm{i} \eta} - \mathrm{e}^{-\mathrm{i} \gamma} \right) & -3 \left( \mathrm{e}^{-\mathrm{i} \eta} +\mathrm{e}^{-\mathrm{i} \gamma} \right)\\
\end{pmatrix}\;.
\end{equation}
Here $\mu=3 k_{x}l , \eta = \left( 3 k_{x} - \sqrt{3} k_{y} \right)l/2$, $\gamma = \left( 3 k_{x} + \sqrt{3} k_{y} \right)l/2$, and  $\varepsilon=c_\varepsilon /c$ is a small non-dimensional parameter.

Non-trivial solutions then follow from the roots of the determinant of the  coefficient matrix in (\ref{EquationsOfMotionMatrixForm1}){, which} yields a polynomial of the sixth order in $\omega^2$. We apply the same normalisations as in Section \ref{SectionHexagonalLattice}, only this time we introduce  $\tilde{\bm C}_\varepsilon = {\bm C}_\varepsilon /c$ (where the "tilde" will again be omitted for convenience in what follows). Next, for $\varepsilon\to 0$, we analyse the behaviour of eigenfrequencies as solutions associated with
\begin{equation}\label{DispersionRelationNormalised1}
\mathrm{det} \left[ {\bm C}_\varepsilon - {\omega}^{2} \left( {\bm I} - {\bm A} \right) \right] = 0 \; .
\end{equation}

\subsection{Dispersive features of the chiral triangular lattice as it degenerates}
\label{SectionDispersionTriangularLattice}

%%%%%%%%%%%%%%%%%%%%%%%%%%%%%%%%%%%%%%%%%%%%
\begin{figure}%[!htcb]
\centering
\includegraphics[width=12cm]{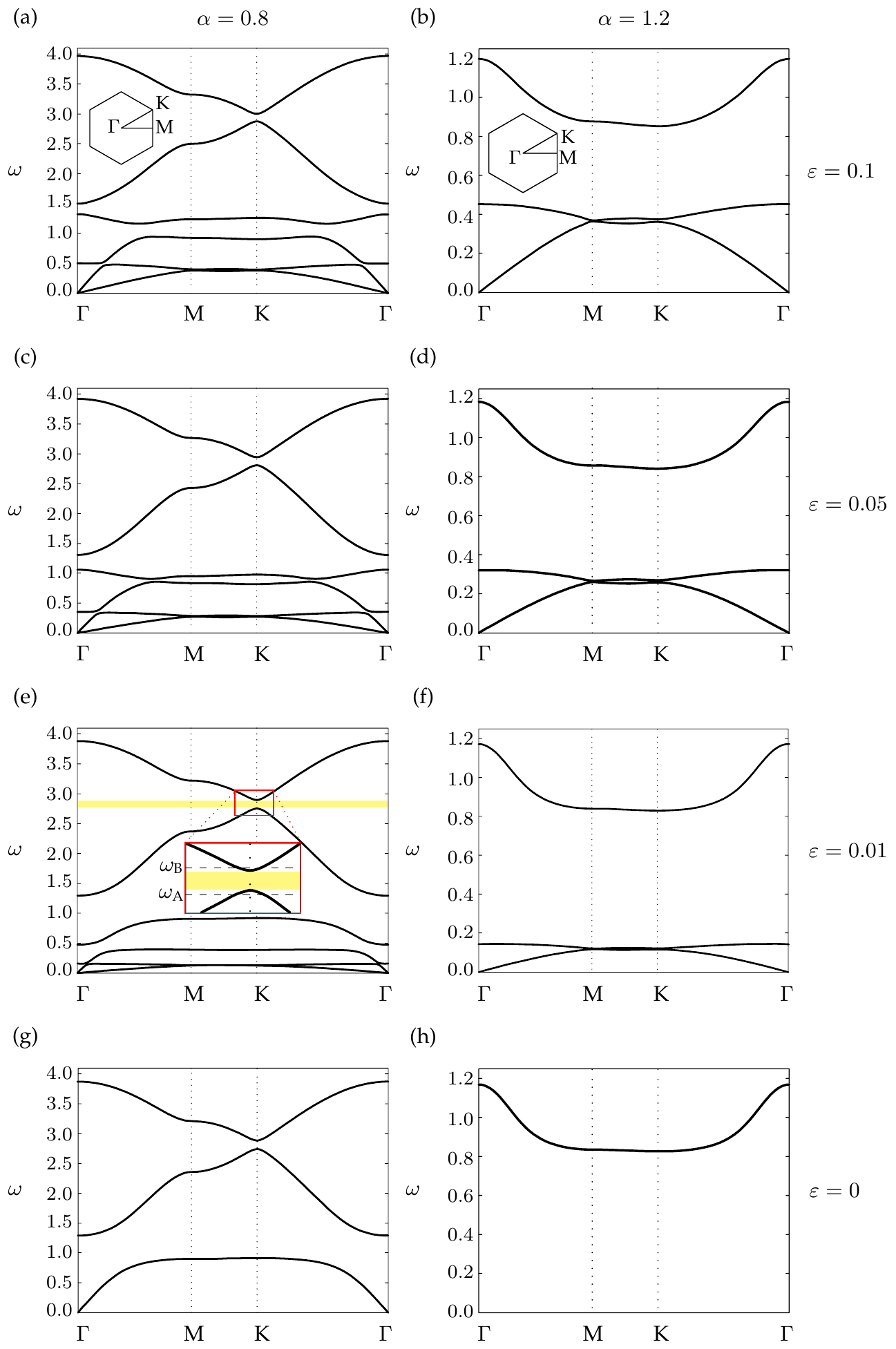}
\caption{\footnotesize Cross-sections of the dispersion surfaces of the chiral triangular lattice, determined for (a, c, e, g) $\alpha = 0.8$ and (b, d, f, h) $\alpha = 1.2$ and for decreasing values of $\varepsilon$: (a,b) $\varepsilon = 0.1$, (c,d) $\varepsilon = 0.05$, (e,f) $\varepsilon = 0.01$, (g,h) $\varepsilon = 0$. The discrete triangular system exhibits six (three) dispersion surfaces for $\alpha < 1$ ($>1$). A magnified inset of the dispersion diagram is included in (e).}
\label{DispersionSurfacesApp}
\end{figure}
%%%%%%%%%%%%%%%%%%%%%%%%%%%%%%%%%%%%%%%%%%%%

Here, we discuss the dispersive nature of the inhomogeneous triangular structure. In particular, we determine the dispersion curves for this system as one traverses the reciprocal lattice space along the same path ${\rm \Upgamma M K\Upgamma}$ described in Section \ref{SectionDispersionHexagonalLattice} for the hexagonal structure.

First, we note that (\ref{DispersionRelationNormalised1}) does not allow for rigid-body motions as observed in the hexagonal lattice in Section \ref{SectionDispersionHexagonalLattice}.  When $0\le \alpha<1$, the relation (\ref{DispersionRelationNormalised1}) admits six positive solutions for $\omega$, namely two acoustic branches and four optical branches. On the other  hand, if $\alpha>1$, only three positive solutions of (\ref{DispersionRelationNormalised1}) exist (the other three solutions for $\omega^2$ are negative).
This can be observed, for instance, in  Fig. \ref{DispersionSurfacesApp}a and Fig. \ref{DispersionSurfacesApp}b, where dispersion curves based on (\ref{DispersionRelationNormalised1}) have been presented for the cases $\alpha=0.8$ and $\alpha=1.2$, respectively; in both figures, the ratio of the stiffnesses of the two types of links is $\varepsilon = 0.1$.
Therefore, the {inhomogeneous} triangular lattice presents three more possible wave modes than the hexagonal structure when $\alpha<1$ and two more when $\alpha>1$ for a given wavevector ({except} at a small number of degenerate points).

For $\alpha<1$, Figs. \ref{DispersionSurfacesApp}a,c,e,g show the dispersion diagram of the triangular lattice as the stiffness $c_\varepsilon$ decreases to zero.
In these figures, we observe that when $\alpha<1$ there are three finite stop-bands.  As $\varepsilon$ decreases, the six dispersion curves move to lower frequencies and the widths of the lowest two finite stop-bands decrease and they disappear when $\varepsilon=0$. This limit results in only three dispersion curves in Fig. \ref{DispersionSurfacesApp}g. These curves represent the non-trivial branches associated with the dispersion relation (\ref{DispersionRelationNormalised}) of the hexagonal lattice connected to gyroscopic spinners, described in detail in Section \ref{SectionDispersionHexagonalLattice} (see also (\ref{EquationsOfMotionDTLa})--(\ref{EquationsOfMotionDTLc}) in connection with this limit).

Similar behaviour is observed for $\alpha>1$ and $\varepsilon \to 0$ in Figs. \ref{DispersionSurfacesApp}b,d,f,h. In this case, one finds three dispersion curves and two finite stop-bands. All curves move to  lower frequencies with $\varepsilon \to 0$.  The lowest two dispersion curves decrease and flatten with decrease of $\varepsilon$ causing the lowest stop-band to shrink. These two curves approach zero and  disappear in passing to the limit as $\varepsilon$ tends to zero. Again, one returns to the case of a hexagonal lattice connected to gyroscopic spinners when $\varepsilon=0$ (see Fig. \ref{DispersionSurfacesApp}h).

%%%%%%%%%%%%%%%%%%%%%%%%%%%%%%%%%%%%%%%%%%%%
\begin{figure}%[!htcb]
\centering
\includegraphics[width=12cm]{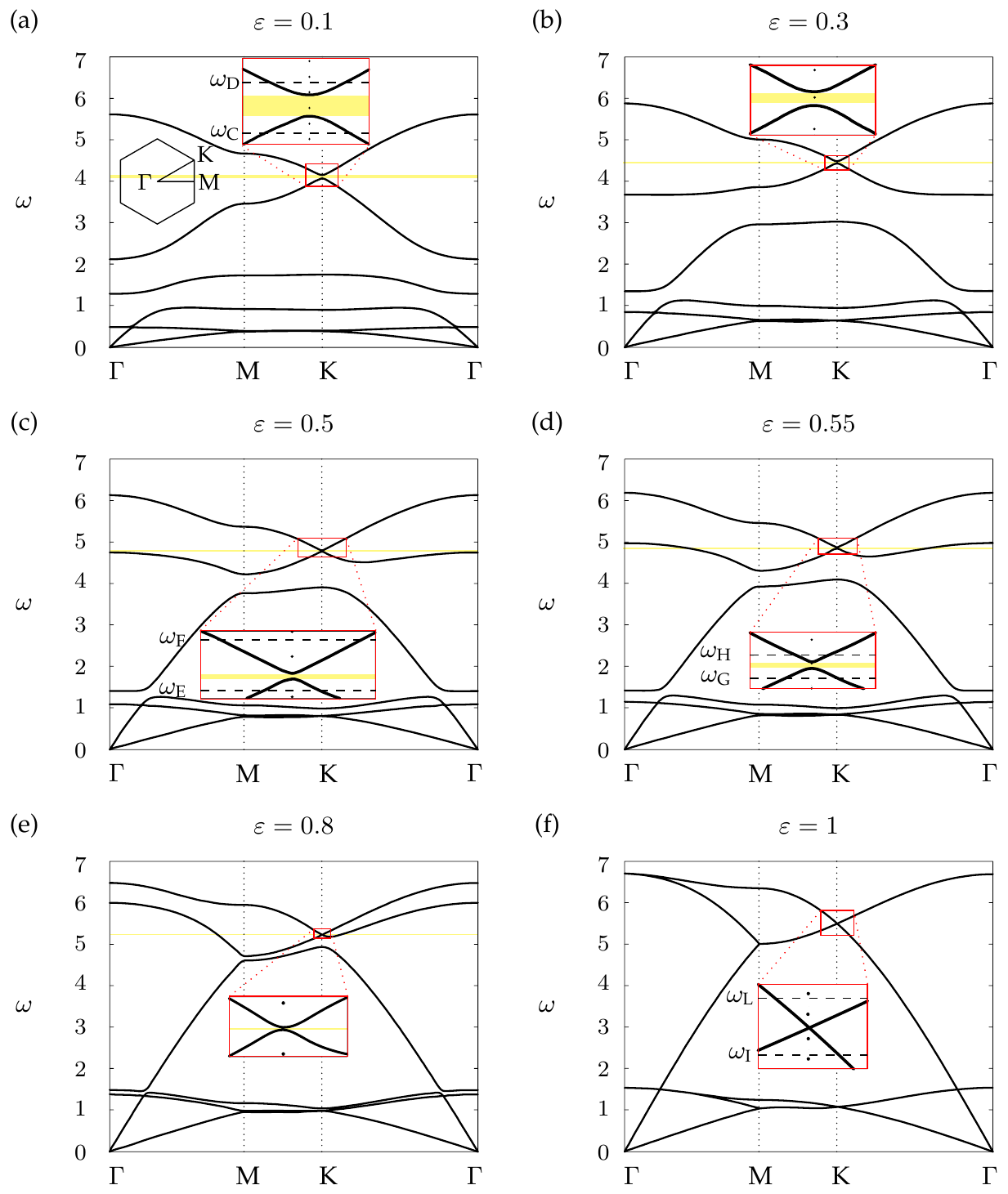}
\caption{\footnotesize Cross-sections of the dispersion surfaces of the chiral triangular lattice, calculated for $\alpha = 0.9$ and for different values of $\varepsilon$, specified in the figures. Magnified insets of the dispersion diagrams are also presented.}
\label{DispersionSurfacesApp2}
\end{figure}
%%%%%%%%%%%%%%%%%%%%%%%%%%%%%%%%%%%%%%%%%%%%

Fig. \ref{DispersionSurfacesApp2} shows the dispersion diagrams for the triangular lattice, determined for a different value of the spinner constant, namely $\alpha = 0.9$. Increasing values for the stiffness of the internal links, represented by $\varepsilon$, are taken. It is apparent that in the low-frequency regime, the effective shear and pressure wave speeds increase with $\varepsilon$, since the system becomes stiffer as $\varepsilon$ is increased. At higher frequencies, the optical branches lift up as the stiffness of the internal links is increased. Moreover, for $\varepsilon \simeq 0.52$ the stop-band in correspondence of the (almost formed) Dirac point becomes a partial stop-band, namely waves can propagate in some directions and are evanescent in the other directions. When $\varepsilon \to 1$, two pairs of dispersion surfaces get closer to each other and they eventually coincide when $\varepsilon = 1$. Fig. \ref{DispersionSurfacesApp2}f represents the dispersion diagram for a homogeneous triangular lattice (see, for instance, \cite{Carta2014}).

\subsection{Uni-directional waveforms in the heterogeneous triangular lattice}
\label{SectionComsolTriangularLattice}

Now, we show that interfacial waveforms with preferential directionality can be also obtained in the heterogeneous triangular lattice. In the numerical simulations presented below, performed in \emph{Comsol Multiphysics}, the lattice domain is a $90 \, l \times 76.2 \, l$ rectangle. The size of the domain is smaller than that used for the hexagonal lattice, since the model for the triangular lattice presents a higher complexity and hence requires a bigger computational cost due to the additional degrees of freedom.

%%%%%%%%%%%%%%%%%%%%%%%%%%%%%%%%%%%%%%%%%%%%
\begin{figure}%[!htcb]
\centering
\includegraphics[width=12cm]{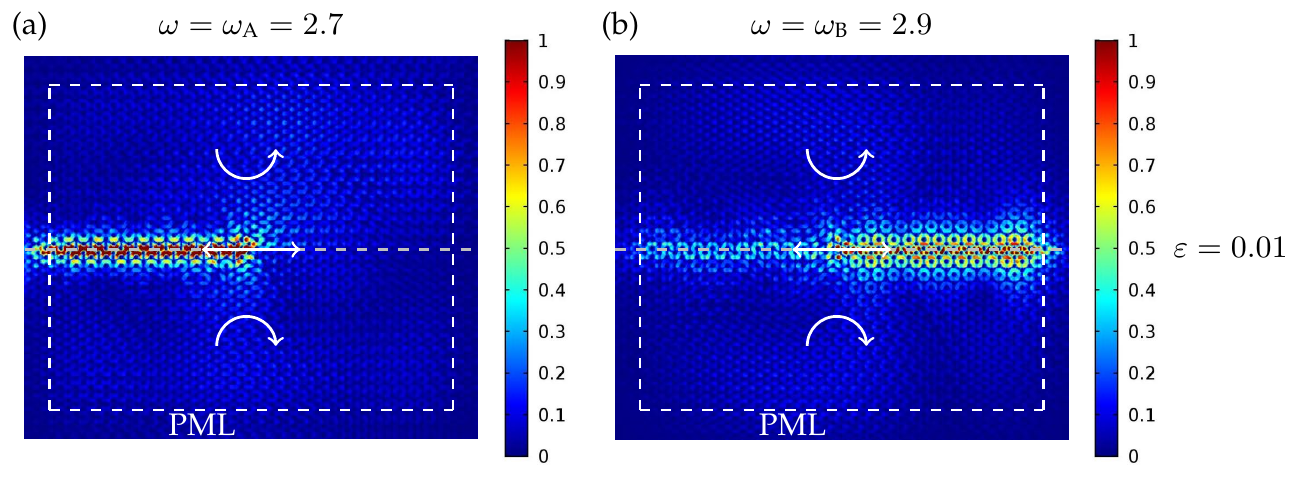}
\caption{Elastic triangular lattice with soft internal links, whose domain is divided into two regions where the spinners rotate in opposite directions. A harmonic displacement is imposed to a node on the interface between the two regions. Interfacial waveforms are generated, whose direction depends on the frequency of the imposed displacement: (a) $\omega = \omega_{\rm A} = 2.7$, (b) $\omega = \omega_{\rm B} = 2.9$ (see also inset in Fig. \ref{DispersionSurfacesApp}e). In the computations, the absolute value of the spinner constant is $\left|\alpha\right|=0.8$ and the links are very soft ($\varepsilon = 0.01$).}
\label{InterfacialWavesApp}
\end{figure}
%%%%%%%%%%%%%%%%%%%%%%%%%%%%%%%%%%%%%%%%%%%%

In Fig. \ref{InterfacialWavesApp}, we divide the lattice domain in two regions, where the gyroscopic spinners are characterised by spinner constants of the same magnitude ($\left|\alpha\right|=0.8$) but of opposite sign. We apply a harmonic displacement of unit amplitude to a lattice particle situated on the horizontal interface between the two regions. In part (a), the radian frequency is $\omega = \omega_{\textup A} = 2.7$, which is below the lower limit of the stop-band highlighted in yellow in Fig. \ref{DispersionSurfacesApp}e, determined from the study of the free vibrations in the analogous homogeneous infinite structure. Conversely, in part (b) the radian frequency of the external excitation is equal to $\omega = \omega_{\textup B} = 2.9$, which lies above the upper limit of the yellow stop-band in Fig. \ref{DispersionSurfacesApp}e.

Comparing Figs. \ref{InterfacialWavesApp}a and \ref{InterfacialWavesApp}b, we observe that uni-directional interfacial waves can be created in the heterogeneous triangular lattice with soft internal links. This is due to the non-trivial topology of the band diagram, characterised by the presence of (almost formed) Dirac points. As in the illustrations relative to the hexagonal lattice studied in Section \ref{SectionComsolHexagonalLattice} and presented in Fig. \ref{InterfacialWaves1}a,b, the wave directionality depends on the frequency of the external excitation. The thickness of the perturbed region in Fig. \ref{InterfacialWavesApp}, where displacements are not zero, is similar to that in Fig. \ref{InterfacialWaves1}, although it seems to be larger due to the smaller size of the domain of the triangular lattice.  We mention that also for the triangular lattice the direction of wave propagation can be reversed by changing the spin directions of the gyroscopic spinners.

%%%%%%%%%%%%%%%%%%%%%%%%%%%%%%%%%%%%%%%%%%%%
\begin{figure}%[!htcb]
\centering
\includegraphics[width=12cm]{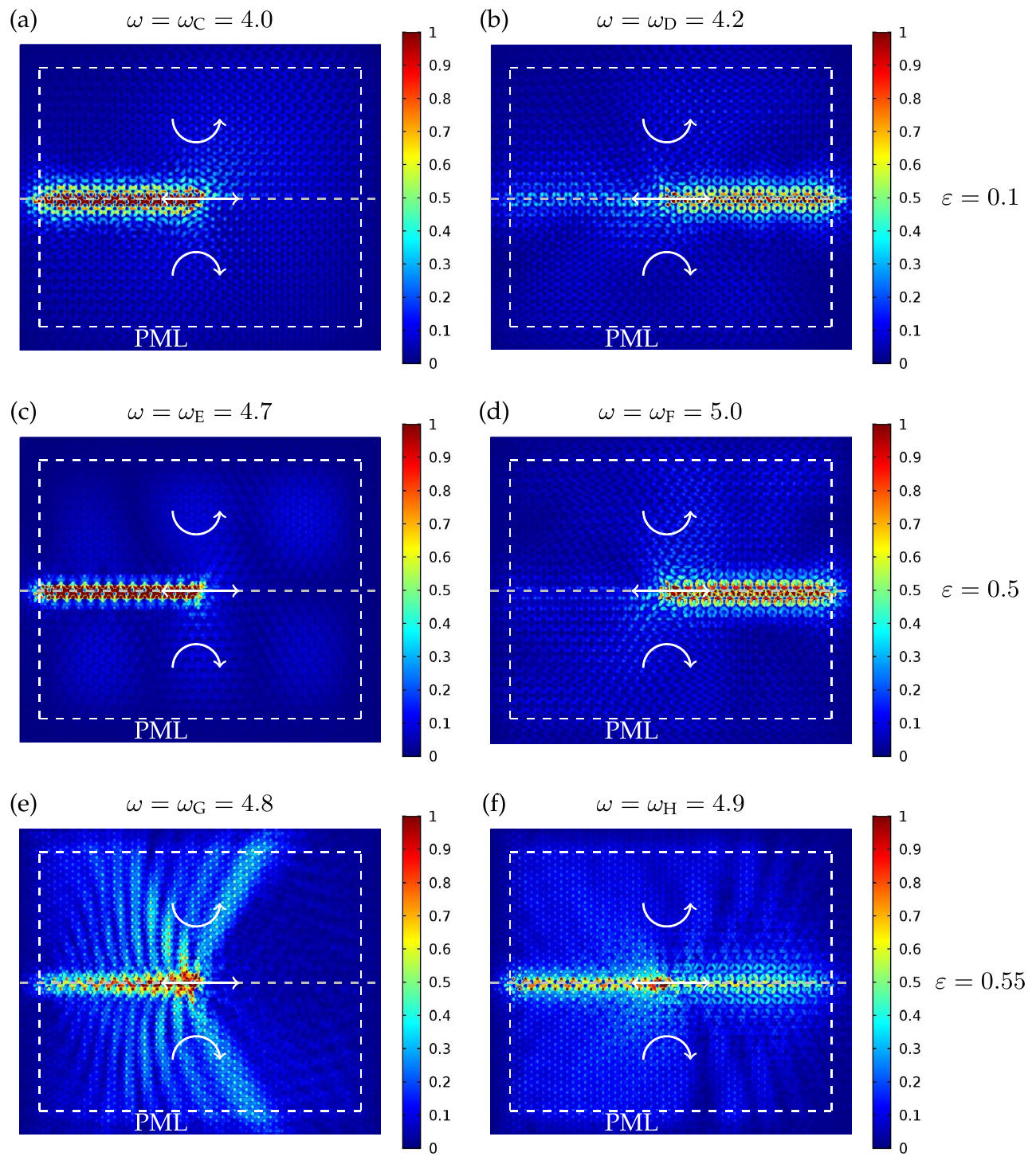}
\caption{Computations showing the response of different triangular lattices.  Each lattice is attached to an inhomogeneous array of spinners. The stiffness of the internal links inside the elementary cell of the triangular lattice is always lower than that possessed by the links on the boundary of the cell. The computational set up is described in Fig. \ref{InterfacialWavesApp}, but  in this case the absolute value of the spinner constant is $\left|\alpha\right|=0.9$. The frequency of the imposed harmonic displacement is (a) $\omega = \omega_{\rm C} = 4.0$, (b) $\omega = \omega_{\rm D} = 4.2$, (c) $\omega = \omega_{\rm E} = 4.7$, (d) $\omega = \omega_{\rm F} = 5.0$, (e) $\omega = \omega_{\rm G} = 4.8$, (f) $\omega = \omega_{\rm H} = 4.9$ (refer to Fig. \ref{DispersionSurfacesApp2} for the positions of the frequencies in the dispersion diagrams). The value of the ratio between the stiffnesses of the links in the elementary cell of the lattice is given by (a,b) $\varepsilon = 0.1$, (c,d) $\varepsilon = 0.5$, (e,f) $\varepsilon = 0.55$.}
\label{InterfacialWavesApp2}
\end{figure}
%%%%%%%%%%%%%%%%%%%%%%%%%%%%%%%%%%%%%%%%%%%%

Fig. \ref{InterfacialWavesApp2} shows the same lattice structure as in Fig. \ref{InterfacialWavesApp}, but when the absolute value of the spinner constant is $\left|\alpha\right|=0.9$. The stiffnesses of the internal links considered in Figs. \ref{InterfacialWavesApp2}a-f are larger than the value chosen in Figs. \ref{InterfacialWavesApp}a,b. Nonetheless, if the system exhibits a stop-band in correspondence of the (almost formed) Dirac point, in particular for $\varepsilon = 0.1$ and $\varepsilon = 0.5$ (see Figs. \ref{DispersionSurfacesApp2}a,c), uni-directional wave propagation can be realised in the triangular lattice (see Figs. \ref{InterfacialWavesApp2}a-d). On the other hand, when the stop-band in the vicinity of the (almost formed) Dirac point is only partial (see Fig. \ref{DispersionSurfacesApp2}d), interfacial waves cannot be generated since waves propagate also in the bulk of the lattice (see Figs. \ref{InterfacialWavesApp2}e,f).

%%%%%%%%%%%%%%%%%%%%%%%%%%%%%%%%%%%%%%%%%%%%
\begin{figure}%[!htcb]
\centering
\includegraphics[width=12cm]{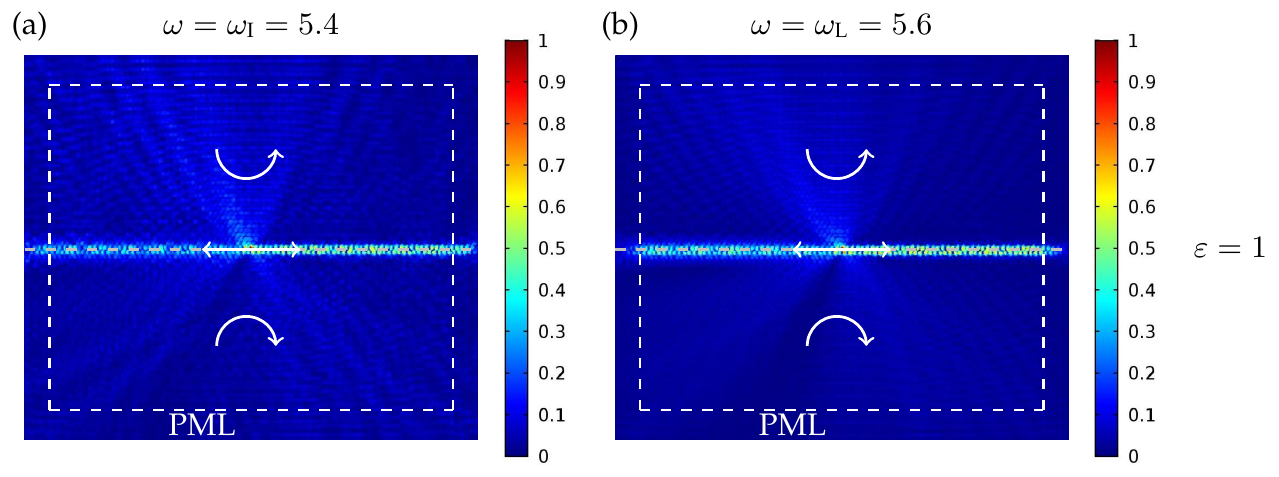}
\caption{Displacement field in the homogeneous triangular lattice ($\varepsilon = 1$) produced by a harmonic displacement, having a frequency (a) lower and (b) higher than the frequency at the Dirac point (see Fig. \ref{DispersionSurfacesApp2}f). In both cases, the absolute value of the spinner constant is $\left|\alpha\right|=0.9$.}
\label{InterfacialWavesApp3}
\end{figure}
%%%%%%%%%%%%%%%%%%%%%%%%%%%%%%%%%%%%%%%%%%%%

Now, we investigate the limit case when $\varepsilon = 1$, that is when the triangular lattice is homogeneous. The corresponding dispersion diagram, plotted in \ref{DispersionSurfacesApp2}f, highlights the disappearance of the yellow stop-band exhibited by the structures with $0 \le \varepsilon < 1$ and the formation of the Dirac point. The response of the homogeneous triangular lattice under a harmonic displacement is shown in Figs. \ref{InterfacialWavesApp3}a and \ref{InterfacialWavesApp3}b when the frequency is $\omega = \omega_{\rm I} = 5.4$ (below the Dirac point) and $\omega = \omega_{\rm L} = 5.6$ (above the Dirac point), respectively. Wave localisation is observed on the interface between the two media where spinners rotate in opposite directions, but without preferential directionality. In addition, waves of small amplitude propagate within the bulk. These results demonstrate that preferential directionality can be realised only if the system exhibits a total stop-bands near the Dirac point.

Finally, we note that interfacial waveforms with hexagonal or rhombic shape, determined for the hexagonal lattice in Section \ref{SectionComsolHexagonalLattice}, can be generated also for the heterogeneous triangular lattice. For the sake of brevity, they will not be presented here.

\section{Conclusions}
\label{Conclusions}
We have developed a novel method of creating topologically protected states in discrete periodic media without perturbing the structured system. This has been achieved by the introduction of gyroscopic spinners, which are linked to individual nodes of the structure.

The full dispersion analysis of a hexagonal lattice connected to gyroscopic spinners has been presented and special features of the dispersive behaviour have been identified. In particular, we have shown that using gyroscopic spinners provides an effective tool in manipulating stop-bands and Dirac points. The latter have then been utilised to create uni-directional interfacial waveforms with a tunable direction. To the best of our knowledge, this represents the first time that  inhomogeneous structured media have been designed to exhibit interfacial waveforms with preferential directionality.

Finite element analysis has been used to demonstrate this phenomenon for a variety of configurations. In certain situations, the distribution of energy along internal boundaries maintains its intensity in the direction of propagation, whereas  along boundaries with discontinuities one can notice a  reduction in the wave amplitude due to scattering brought by the discontinuities.

The hexagonal lattice represents the limit case of a triangular lattice with very soft internal links within its hexagonal cells. Indeed, when gyroscopic spinners are attached to the triangular lattice, interfacial waves can be generated at the boundaries of regions where spinners rotate in opposite directions. The directionality of the interfacial waveforms can be swapped by inverting the spin directions of the gyroscopic spinners or by changing the frequency of the excitation.

We anticipate that the designs proposed here will lead to new ways of generating and controlling topologically protected states in discrete media through the use of chiral metamaterials.

%\contr{%
%M.G. performed the analytical and numerical computations for the work.
%All the authors contributed to the development of the model and the preparation of the manuscript.
%{\color{red}Write contributions of the authors.}
%}

%\disclaimer{Insert disclaimer text here.}

%\ethics{Insert ethics text here.}

%\ack{...}

{{\bf Acknowledgements:}
%{\color{red}M.G. wishes to thank the support of ... .}
 M.J.N. gratefully acknowledges the support of the EU H2020 grant MSCA-IF-2016-747334-CAT-FFLAP. G.C., I.S.J., N.V.M. and A.B.M. would like to thank the EPSRC (UK) for its support through Programme Grant no. EP/L024926/1.}

%\conflict{The authors have no conflict of interests to declare.}

%\dataccess{This paper contains no experimental data. All computational results are reproducible.}

\end{document}